\documentstyle[prl,aps,epsf]{revtex}
\begin{document}
\draft
\title{
\large \bf
Optical Susceptibilities of Polymers:\\
Current-Current versus Dipole-Dipole Correlation
}
\author{Minzhong Xu}
\address{
Department of Chemistry, New York University, New York, NY 10003, U.S.A.
        }
\author{Xin Sun}
\address{ 
Department of Physics, Fudan University, and National Laboratory of Infrared
Physics, Shanghai 200433, P.R. China
}
\date{\today}
\maketitle
\bigskip
\begin{abstract}
The static current operator leads to definitional zero frequency divergence
and unphysical results in studying nonlinear optical susceptibilities of 
polymers. A well-defined dipole-dipole correlation is superior to the 
complicated current-current correlation to solve this problem. As illustrative 
examples, optical susceptibilities under both SSH and TLM models of 
trans-(CH)$_x$ are studied. New analytical results are obtained. The 
reasons of previous improper results are analyzed.
\end{abstract}
\phantom{.}

{To study the nonlinear optical (NLO) properties of polymers, periodic 
approximate models are necessary to simplify the real systems. Some good 
approximate models of polymers based on the tight-binding 
approximation (TBA), such as SSH\cite{ssh} and TLM\cite{tlm} models in weakly 
correlated systems, Hubbard and Pariser-Parr-Pople models in strongly 
correlated and electron-electron ($e$-$e$) interaction systems, etc, have 
yielded physical insights surpassing the complicated non-approximate
computations. In considering the optical response of these models, a U(1)
transformation has been suggested\cite{fradkin,gebhard,ferry} to provide the 
gauge invariance of the TBA Hamiltonian. 

In linear optical (LO) response theory, the Kubo formula based on 
current-current ($J$-$J$) correlation\cite{martin,mahan} is widely used because 
of simplicity. It is commonly held that in discussing the optical 
susceptibilities of materials, the $J$-$J$ correlation (${\bf p \cdot A}$) will 
play the same role as that of dipole-dipole ($D$-$D$) correlation (${\bf E 
\cdot r}$)\cite{shen,shaul0} and that the apparent definitional zero frequency 
divergence (ZFD) in the NLO susceptibilities definition\cite{butcher,wwu} is 
only a virtual problem, although the proofs only have been shown under certain 
assumptions.\cite{martin,aspnes} However, there exists some discrepancies from 
the above conclusion under the models: (i) In the study of conductivity $\sigma$
under TBA models,\cite{bishop,maki} the $J$-$J$ correlation will have ZFD in 
Im($\sigma$) (related to the real part of $J$-$J$) although it can give the 
correct Re($\sigma$) (related to the imaginary part of $J$-$J$). (ii) In 
third-harmonic generation (THG) of trans-(CH)$_x$, the experimentally observed
two-photon absorption peak (TP)\cite{kajzar} has given rise to the theoretical 
explanations. Based on $J$-$J$ correlation, the TP was explained\cite{wwu} 
by choosing the TLM model. But it has been criticized by 
others\cite{su,cwu,shuai,mazumdar,soos} with simple parity consideration.
Also the spectrum\cite{wwu} is quite different from that under the $D$-$D$ 
correlation.\cite{mxu} Thus, it casts some doubts in the practical application 
of $J$-$J$ correlation. 

In this letter, we will show that all above controversies are caused by 
the improper application of $J$-$J$ correlation 
based {\it only} on the static current operator $\hat{J_0}$. To recover the 
correct results, a well-defined $D$-$D$ correlation is more suitable for 
studying the optical susceptibilities than the $J$-$J$ correlation, whose 
equivalence to the $D$-$D$ correlation can be satisfied by introducing 
complicated induced field currents (IFCs). Besides solving the apparent ZFD in 
the definition of NLO susceptibility, the $D$-$D$ correlation is to be favored
over the $J$-$J$ correlation due to the lack of the gauge dependence of the 
vector potential ${\bf A}$\cite{shaul0} and simplifying the definitional 
complexity of the contribution by IFCs in the high order 
expansions.\cite{fradkin,gebhard} As a deduction, the TP cusp and 
ZFD\cite{wwu,bishop,maki} will no longer exist in both SSH and TLM models by 
the well-defined $D$-$D$ correlation. However, this seemingly trivial conclusion
has not been clearly illustrated by the others although the $D$-$D$ correlation 
has already been applied in the NLO response of the real 
systems.\cite{shaul0,su,soos}

The application of $J$-$J$ correlation is simplier than that of the $D$-$D$ 
correlation in LO response because of the convenient search for the static 
current $\hat{J_0}$\cite{mahan} compared with that for dipole expression 
$\hat{D}$ under the approximate models. As a typical example, the position 
operator $\hat{\bf r}$ is ill-defined in periodic systems\cite{aspnes,callaway}
in real space while $\hat{J_0}$ is not. Further, the static 
current $\hat{J_0}$ still can give the correct results in the LO 
absorption\cite{mahan,bishop,maki} because the IFCs only contribute to the real 
part of $J$-$J$ correlation. Thus the simplicity of the Kubo formula is still 
satisfied by applying the static current $\hat{J_0}$ and ZFD in the real part 
of $J$-$J$ correlation could be solved either by the Kramers-Kronig (KK) 
relation\cite{mahan} or by subtracting  
$\langle [j, j] \rangle(\omega=0)$.\cite{bishop} It is not a big 
surprise to see why the correct results still can be preserved for the LO 
response through $\hat{J_0}$. However, the advantage of $J$-$J$ correlation 
application in LO response is no longer available for the NLO response because 
of the IFCs,\cite{gebhard,butcher} thus the definition of $n$-th order $J$-$J$ 
correlation based {\it only} on the static current\cite{wwu} will lead to the 
incorrect results. Besides the difficulty in obtaining the correct IFCs in 
approximate models, $J$-$J$ correlation will make the definition of NLO 
susceptibilities unable to be written in a general form as defined in Wu' work 
\cite{wwu} and will make the computations very tedious,\cite{gebhard}
although it may give the correct result.

In periodic systems, the imposed boundary condition requires (i) Bloch wave 
functions to describe the extended electronic states and (ii) all operators 
including the position operator $\hat{\bf r}$ to have at least the same 
periodicity.\cite{weinreich} A band index n and crystal momentum {\bf k} are
used to label the Bloch states: $\displaystyle |n, {\bf k}> = u_{n, {\bf k}} 
({\bf r})e^{i{\bf k \cdot r}}$, where $u_{n, {\bf k}}({\bf r})$ is the periodic
function under the translation of lattice vector. If the homogeneous electric
field is applied, the energy level $E_n({\bf k})$ should be changed to
$E_n({\bf \kappa})$\cite{ferry} and the static current operator 
$\hat{J}_0({\bf k})$ should be replaced by a new current operator 
$\hat{J}({\bf \kappa})$, where 
${\bf \kappa}={\bf k}-e{\bf A}/\hbar$.\cite{butcher} By Taylor expansion of 
$\hat{J}({\bf \kappa})$ based on powers of the vector potential ${\bf A}$, we
will obtain  the IFCs related to all orders of ${\bf A}$ besides the static 
current $\hat{J_0}$. There is no doubt that the IFCs will contribute
to the optical response. Complexities in handling IFCs (usually 
related to the intra-band currents\cite{butcher}) in NLO studies could be 
imagined. Polymer models\cite{ssh,tlm} usually share the same properties as
periodic systems. The direct consequence of dropping the IFCs will result in 
ZFD and unphysical results.\cite{wwu,bishop,maki} Although the unphysical 
results in NLO response have been recurrently questioned by the 
others,\cite{su,cwu,shuai,mazumdar,soos} the direct reason has never been 
revealed and correct analytical results have never been obtained. 
The chief difficulty is that {\it direct} ways to include IFCs in many
approximate models are forbidden\cite{ssh,tlm,bishop} and the static current 
could easily be improperly used.\cite{wwu,maki} By a well-defined $D$-$D$
correlation, we will show that all difficulties caused by the improper 
use of $J$-$J$ correlation could be solved. To give an intuitive 
picture, we will illustrate those correlations by studying the simple 
one-dimensional ($1$-$d$) TBA electron-lattice models of trans-(CH)$_x$ (SSH 
and TLM) whose optical properties have been widely studied by
others.\cite{wwu,maki,kajzar,su,cwu,shuai,mazumdar,soos,heeger} To avoid the 
ill-definition and to provide the periodicity of the position operator 
$\hat{\bf r}$, we express $\hat{\bf r}$ under $|n, {\bf k}>$ as:\cite{callaway}
\begin{eqnarray}
{\bf r}_{n {\bf k}, n' {\bf k'}}= i \delta_{n,n'}{\bf \nabla_{k}}
\delta({\bf k}-{\bf k'}) + \Omega_{n,n'}({\bf k})\delta({\bf k}-{\bf k'}), 
\label{r}
\end{eqnarray}
where $\displaystyle \Omega_{n,n'}({\bf k})=\frac{i}{v}\int_{v}
u_{n,{\bf k}}^*({\bf r}){\bf \nabla_{k}} u_{n', {\bf k}}({\bf r}) d {\bf r}$
and $v$ is unit cell volume.

{\it Susceptibilities definition by $D$-$D$ correlation:} Without considering 
retardation effect,\cite{shaul0} optical susceptibilities are usually defined
by expanding the optical polarization in powers of the transverse electric 
field.\cite{shen} Under that approximation, the general $n$th-order 
susceptibility is a purely material quantity defined as:\cite{shaul0}

\begin{eqnarray}
\chi^{(n)}(\Omega; \omega_{1}, \ldots, \omega_{n})= \frac{1}{n!}
\left[ \frac{i}{\hbar} \right]^n \int  d{\bf r}_{1} \cdots d{\bf r}_{n}
\int dt_{1} \cdots dt_{n}
\int d{\bf r} dt\, e^{-i {\bf k \cdot r}+ i \Omega t} \langle \hat{T}
\hat{{\bf D}} ({\bf r},t) \hat{{\bf D}}({\bf r}_{1},t_{1}) \cdots
\hat{{\bf D}} ({\bf r}_{n},t_{n}) \rangle,
\label{DD}
\end{eqnarray}
where $\displaystyle \Omega \equiv -\sum_{i=1}^{n} \omega_{i} $,
$\hat{T}$ is the time-ordering operator and $\hat{\bf D}$ is dipole operator.

Based on periodic TBA, SSH Hamiltonian\cite{ssh} is given:
\begin{eqnarray}
H_{SSH}=-\sum_{l,s} \left[ t_0+(-1)^l \frac{\Delta}{2} \right]
(\hat{C}_{l+1,s}^{\dag}\hat{C}_{l,s}^{}+\hat{C}_{l,s}^{\dag}\hat{C}_{l+1,s})^{},
\nonumber
\end{eqnarray}
where $t_0$ is the transfer integral between the nearest-neighbor sites,
$\Delta$ is the gap parameter and $\hat{C}_{l,s}^{\dag}(\hat{C}_{l,s})$
creates(annihilates) an $\pi$ electron at site $l$ with spin $s$. In
continuum limitation, above SSH model will give the TLM model.\cite{tlm}

In momentum space, the above Hamiltonian with electron-photon ($e$-$\bf{A}$) 
interaction could be found as follows:
\begin{eqnarray}
H(k,t)= \sum_{k,s} \varepsilon(k) \hat{\psi}_{k,s}^{\dag}(t) \sigma_{3}
\hat{\psi}_{k,s}(t)- \hat{D} \cdot E e^{i\omega t},
\label{H}
\end{eqnarray}
where $\varepsilon (k)$$=$$\sqrt{\left[ 2 t_0 cos(ka) \right]^2+\left[ \Delta
sin(ka) \right]^2}$ and $\hat{\psi}_{k,s}^{\dag}(t)$=
$(\hat{a}^{\dag c}_{k,s}(t)$, $\hat{a}^{\dag v}_{k,s}(t))$ is the two-component 
spinor describing excitations of electrons in the conduction band and valence 
band. Long wave approximation\cite{mahan} is applied in electromagnetic field 
${\bf E}$ with frequency $\omega$. The dipole operator $\hat{D}$ could be 
obtained by the Eq. (\ref{r}):

\begin{eqnarray}
\hat{D}= e \sum_{k,s}(\beta(k)\, \hat{\psi}_{k,s}^{\dag}
\sigma_{2}\hat{\psi}_{k,s}
 +i \frac{\partial}{\partial k} \, \hat{\psi}_{k,s}^{\dag}\hat{\psi}_{k,s}),
\label{D}
\end{eqnarray}
where $\beta(k)=-\Delta t_0 a / \varepsilon^2(k)$, is the 
coefficient related to the interband transition between the conduction 
and valence bands in a unit cell $2a$ and the second term is related to
the intraband transition,\cite{aspnes} $e$ is the electric charge and 
$\vec{\sigma}$ are the Pauli matrixes. We neglect the relative distortion
$\eta$($\equiv 2u/a$) in the dipole operator because it is relatively
small in the optical contribution.\cite{gebhard,mxu}

Due to the fact that $\pi$ electrons in the SSH model are 
non-localized,\cite{heeger} the dipole approximation\cite{shaul0,soos} is no 
longer valid in the extended states and will lead to wrong results as pointed 
out by some authors.\cite{gebhard} The Fourier transform of Eq. (\ref{D})
to coordinate space shows that the transition dipole is related to the electron 
hopping to {\it all} the other sites besides the nearest neighbor sites. Thus 
the dipole approximation fails for the extended states in periodic systems. 
Because of the failure of dipole approximation through the polarization operator
$\hat{P}$, computations show a magnitude difference of $10^2$ in $\chi^{(1)}$
and $10^4$ in $\chi^{(3)}$\cite{mxu} and quite different shape in spectrum 
compared with the results through the dipole operator $\hat{D}$ and the 
experimental values\cite{kajzar,heeger} in trans-(CH)$_x$, although
the position of some resonant peaks may be correctly obtained. 

{\it LO response by $D$-$D$:} The LO susceptibility 
$\chi_{SSH}^{(1)}(\Omega, \omega_1)$ can be obtained from Eq. (\ref{DD}) and 
Eq. (\ref{D}):
\begin{eqnarray}
\chi_{SSH}^{(1)}(-\omega_1, \omega_1)=2\left[ \frac{i}{\hbar} \right]
e^2 \sum_{k} \int_{-\infty}^{\infty} &T&r \Biggl\{
i\frac{\partial}{\partial k} \left[G(k,\omega) i \frac{\partial}{\partial k}
\left[G(k,\omega -\omega_1) \right]\right]
+\beta(k) \sigma_{2} G(k,\omega) i \frac{\partial}{\partial k} \left[
G(k,\omega-\omega_1) \right] \nonumber\\
&+&i \frac{\partial}{\partial k} \left[ \beta(k) G(k,\omega) \sigma_2
G(k,\omega-\omega_1) \right]
+\beta(k) \sigma_2 G(k,\omega) \beta(k) \sigma_2 G(k,\omega-\omega_1)
\Biggr\} \frac{d \omega}{2 \pi},
\label{ssh1}
\end{eqnarray}
where the Green function $ \displaystyle G(k,\omega)=\frac{\omega+\omega_k 
\sigma_3}{\omega^2-\omega^2_k+i \epsilon} \, \, $ with $\omega_k \equiv 
\varepsilon(k) / \hbar \text{ and } \epsilon \equiv 0^+$.

By Eq. (\ref{ssh1}), we have $\chi_{SSH}^{(1)}(\omega)$$\equiv$
$\chi_{SSH}^{(1)}(-\omega,\omega)$:
\begin{eqnarray}
\chi_{SSH}^{(1)}(\omega)
= \frac{e^2(2 t_0 a)}{2 \pi \Delta^2} \int_{1}^{\frac{1}{\delta}}
\frac{d x}{ [(1-\delta^2 x^2)(x^2-1)]^{\frac{1}{2}} x^2 (x^2-z^2)},\nonumber
\end{eqnarray}
where $ x \equiv \hbar \omega_k / \Delta$, $z \equiv \hbar \omega /(2\Delta)$
and $\delta \equiv \Delta /(2 t_0)$.

If the continuum limitation is applied, that is, $\delta \to 0 \text{ and }
2 t_0 a \to \hbar v_F $, the above intergral gives the LO susceptibility 
$\chi_{TLM}^{(1)}(\omega)$ under the TLM model\cite{tlm} as follows:
\begin{eqnarray}
\chi_{TLM}^{(1)}(\omega)=-\frac{e^2\hbar v_F}{2 \pi \Delta^2 z^2}(1-f(z)),
\label{x1} 
\end{eqnarray}
where
\begin{equation}
f(z) \equiv \left \{
\begin{array}{lr}
\displaystyle  {\arcsin (z)\over z \sqrt{1-z^2}}  &(z^2<1),\\
\\
\displaystyle  -{\cosh^{-1} (z)\over z\sqrt{z^2-1}}+\displaystyle
{i\pi \over 2 z\sqrt{z^2-1}} &\ \ (z^2>1).
\end{array}
\right.
\label{fz}
\end{equation}

The conductivity $\sigma(\omega)$ given by $-i\omega \cdot \chi^{(1)}$, is 
exactly the same as based on $J$-$J$ correlation.\cite{maki} However, we should
point out that the {\it direct} computation based on the static current from 
$J$-$J$ correlation under TLM model\cite{tlm,maki} shows {\it no} first term in
Eq. (\ref{x1}),\cite{mxu} ZFD in real part of $J$-$J$ correlation is 
obvious although the correct imaginary part still can be given. These 
difficulties have never been clearly addressed previously, provided the reason 
that the static current $\hat{J_0}$ is still valid for obtaining the correct 
imaginary part in the LO absorption. To include the IFCs by changing
${\bf k}\to{\bf k}-e{\bf A}/\hbar$ in  the static current operator 
$\hat{J_0}({\bf k})$, ZFD could be solved to give the same result as that 
under $D$-$D$ correlation.\cite{mxu1} But the complicated way to include the
IFCs compared with simple $D$-$D$ correlation already makes $J$-$J$ correlation
impractical even for the LO response. Fortunately, the IFCs only contribute for 
the real part of $J$-$J$ correlation and have no influence on the absorption. 
Attempts to obtain IFCs directly from TLM Hamiltonian fail because it is 
forbidden to include $A^2$ term by the model.\cite{tlm} In the discrete SSH 
Hamiltonian, we have the chance to include the IFCs through Peierls 
substitution,\cite{gebhard} but a straightforward computation easily shows that
they are not correct IFCs to cancel the ZFD.\cite{mxu1} It gives us an 
impression of the difficulty to obtain the correct IFCs besides the complexity 
of handling their contributions to the optical response even in well-defined 
$1$-$d$ periodic models. From the above examples, the feasibility of $J$-$J$ 
correlation in more general models will be questioned and the application of 
$D$-$D$ correlation is more reasonable under approximate models to obtain the 
correct results.

{\it NLO susceptibilities of trans-(CH)$_x$ chain:}
There are many elegant works in discussing NLO susceptibilities of 
polymer chains.\cite{wwu,kajzar,su,cwu,shuai,mazumdar,soos} The 
TP\cite{wwu} obtained from $J$-$J$ correlation was doubted in the 
literature\cite{kajzar,su,cwu,shuai,mazumdar,soos} since it is forbidden by 
momentum conservation and parity consideration in both TLM and SSH models.
Based on the $D$-$D$ correlation, our analytical results of THG show explicitly
that TP {\it no longer} exists in both the SSH and the TLM models. This 
unphysical TP is caused by the same reason -- the improper use of the static 
current $\hat{J}_0$ and omission of the IFCs in periodic chain. Although the 
reason for $J$-$J$ correlation is very clear, we only give the computational 
results based on $D$-$D$ correlation. It is expected that if the correct IFCs 
are considered, $J$-$J$ correlation will give the same results as $D$-$D$ 
correlation although much more complicated computations are inevitable.
After a similar definition as Eq. (\ref{DD}) and tedious derivations, the new 
result of THG per unit length under SSH model for infinite chains is 
recovered as:
\begin{eqnarray}
\chi_{SSH}^{THG}(\omega)
=\chi_0^{(3)} \frac{45}{128} \int_1^{\frac{1}{\delta}}
\frac{d x}{ [(1-\delta^2 x^2)(x^2-1)]^{\frac{1}{2}}}
\Biggl\{
&-&\frac{47-48(1+\delta^2)x^2+48\delta^2x^4}{8 x^8(x^2-z^2)}
+\frac{3(1-\delta^2 x^2)(x^2-1)}{x^6 (x^2-z^2)^2} \nonumber \\
&+&\frac{9\left[ 47-48(1+\delta^2)x^2+48\delta^2 x^4 \right]}
{8 x^8(x^2-(3 z)^2)}+\frac{63(1-\delta^2 x^2)(x^2-1)}
{x^6 (x^2-(3 z)^2)^2} \Biggr\}
\label{ssh3}
\end{eqnarray}
where $\displaystyle \chi_0^{(3)} \equiv \frac{8}{45}\frac{e^4 n_0}{\pi}
\frac{(2 t_0 a)^3}{\Delta^6}$, $n_0$ is the number of chains in unit cross 
area, the polymer chains are assumed to be oriented. $x, \ z \text{ and } 
\delta$ are defined the same as in $\chi^{(1)}_{SSH}(\omega)$.

Eq. (\ref{ssh3}) is an elliptical integration and can be numerically 
integrated if one change $x \to x+i\epsilon$ in considering the life-time of 
the state.\cite{su,cwu} For polyacetylene, by choosing $t_0=2.5 eV$, 
$\Delta=0.9 eV$, $n_0=3.2 \times 10^{14} cm^{-2}$, $a=1.22 \AA$ and 
$\epsilon \sim 0.03$,\cite{cwu} we have $\delta=0.18$ and
 $\chi_0^{(3)} \approx 1.0 \times 10^{-10}$ esu. The absolute value of 
$\chi_{SSH}^{THG}(\omega)$ is plotted in Fig. 1. It shows the good agreement 
with the experimental value\cite{kajzar} around $z=1/3$. 

\begin{figure}
\vskip -10pt
\centerline{
\epsfxsize=7cm \epsfbox{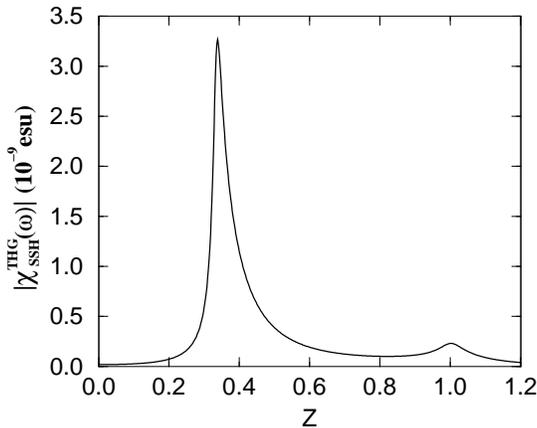}
}
\vskip 0pt
\caption{$|\chi_{SSH}^{THG}(\omega)|$ for $\epsilon$=$0.03$ with
$z \equiv \hbar\omega/(2\Delta)$.}
\end{figure}

\begin{figure}
\vskip -10pt
\centerline{
\epsfxsize=7cm \epsfbox{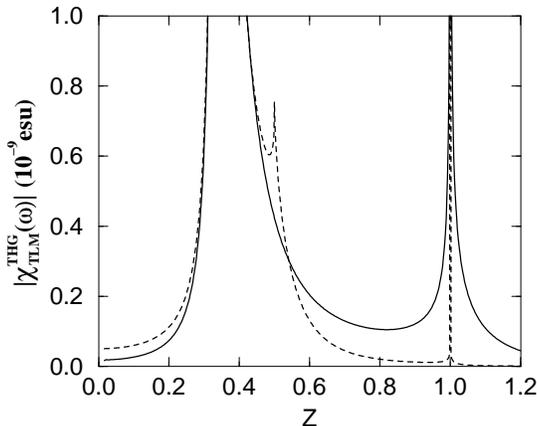}
}
\vskip -10pt
\caption{Computed $D$-$D$ value (solid line) vs. $J_0$-$J_0$ value 
(dashed line)of $|\chi_{TLM}^{THG}(\omega)|$ with 
$z \equiv \hbar\omega/(2\Delta)$.}
\end{figure}

Let $ \delta \to 0^+$ and $\epsilon \to 0^+$ in Eq. (\ref{ssh3}), we
obtain the analytical result of THG under TLM model as follows:
\begin{eqnarray}
\chi_{TLM}^{THG}(\omega)= \chi_0^{(3)} \frac{45}{128} \, \Biggl\{
-\frac{14}{3 z^8}-\frac{4}{15 z^4}
+\frac{(37-24 z^2)}{8 z^8} f(z)+\frac{(1-8 z^2)}{24 z^8} f(3z) \Biggr\}
\label{tlm3}
\end{eqnarray}
where $f(z)$ and $\chi_0^{(3)}$ defined in Eq. (\ref{fz}) and Eq.
(\ref{ssh3}). The comparison between our result($D$-$D$) and Wu's 
result($J_0$-$J_0$)\cite{wwu} on absolute value of $\chi_{TLM}^{THG}$ is
plotted in Fig.2.

The TP disappears in our analytical results, which is more reasonable for the 
physical situation and consistent with the previous numerial 
computations.\cite{su,cwu,shuai,mazumdar,soos} It 
also implies that the $e$-$e$ interactions,\cite{mazumdar,soos} disorders, 
quantum fluctuations or finite chain size effects\cite{cwu} should be taken 
into account to explain this experimentally observed TP.\cite{kajzar}
Fig.1 shows another new resonant peak $z=1$ at ratio of 1/10 of $z=1/3$. 
This new peak hasn't been reported by the experiments\cite{kajzar} because of 
the experimental scanning range. The cancellation of ZFD by $J_0$-$J_0$ 
correlation\cite{wwu} is actually a coincidence under TLM model, because in the
SSH model, we find ZFD in $\chi^{(3)}_{SSH}$ through $J_0$-$J_0$ 
correlation.\cite{mxu} Keldysh formalism\cite{wwu} is not necessary to apply 
in this equilibrium system.\cite{su} From THG of trans-(CH)$_x$, $D$-$D$ 
correlation is superior to $J$-$J$ correlation to obtained NLO susceptibilities.

As a conclusion, the principle about the equivalence of $D$-$D$
vs. $J$-$J$ correlation is still correct under the approximate models
while practicality favors $D$-$D$ over $J$-$J$ correlation. Although our
discussion is chiefly based on $1$-$d$ polymer chains, the main results
in this letter can be generalized for $2$-$d$ or $3$-$d$ periodic systems.

This work was supported by Chemistry Department, New York University, 
the Project $863$ and the National Natural Science Foundation of China
(59790050, 19874014). Very helpful discussions with Z. Ba\v{c}i\'{c}, J.L. 
Birman, H.L. Cui, A.J. Epstein, G. Gumbs, N. H\"{o}ring, Y.R. Shen, Z.G. Shuai,
Z.G. Soos, M.E. Tuckerman, Z. Vardeny, C.Q. Wu, Z.G. Yu and J.Z.H. Zhang are 
acknowledged.

}

\begin{references}
\bibitem{ssh} W.P. Su, J.R. Schrieffer, and A.J. Heeger, Phys. Rev. Lett. 
{\bf 42}, (1979) 1698; Phys. Rev. B {\bf 22}, (1980) 2099.
\bibitem{tlm}Lagrangian: $L=\int dx\hat{\psi}^{\dagger}(x) \{i\hbar \partial_t
+i\sigma_3 v_F \partial_x + \sigma_1\Delta \}\hat{\psi}(x)$. [H. Takayama, 
Y.R. Lin-Liu, and K. Maki, Phys. Rev. B {\bf 21}, (1980) 2388.]; The vector 
potential $A$ is included by changing 
$-i\hbar\partial_x\to-i\hbar\partial_x-eA$, it will not contain $A^2$ term
in the Lagrangian, also see Wu's work.\cite{wwu}
\bibitem{fradkin}E. Fradkin, Field Theories of Condensed Matter Systems
(Addison-Wesley Publishing Company, 1991), p.9.
\bibitem{gebhard}F. Gebhard, K. Bott, M. Scheidler, P. Thomas, and S.W. Koch,
Phil. Mag. B, Vol. {\bf 75}, No.1, (1997) 1-12.
\bibitem{ferry} Quantum Transport in Semiconductors, edited by D.K. Ferry
and C. Jacoboni (Plenum, New York, 1992), p.57.
\bibitem{martin}P. Martin and J. Schwinger, Phys. Rev. {\bf 115}, (1959) 1342.
\bibitem{mahan}G.D. Mahan, Many-Particle Physics (Plenum Press, New York 
$\&$ London, 1981), chap. 3.
\bibitem{shen}Y.R. Shen, The Principles of Nonlinear Optics (John Wiley 
$\&$ Sons, Inc, 1984). 
\bibitem{shaul0}S. Mukamel, Principles of Nonlinear Optical 
Spectroscopy (Oxford, New York, 1995), and references therein.
\bibitem{butcher}P.N. Butcher and T.P. McLean, Proc. Phys. Soc. (London) 
{\bf 81}, (1963) 219; {\bf 83}, (1964) 579.
\bibitem{wwu}Weikang Wu, Phys. Rev. Lett. {\bf 61}, (1988) 1119.
\bibitem{aspnes}D.E. Aspnes, Phys. Rev. B {\bf 6}, (1972) 4648.
\bibitem{bishop}I. Batistic and A.R. Bishop, Phys. Rev. B {\bf 45}, (1992) 5282.
\bibitem{maki}K. Maki and M. Nakahawa, Phys. Rev. B {\bf 23}, (1981) 5005.
\bibitem{kajzar}W.S. Fann, S. Benson, J.M.J. Madey, S. Etemad, G.L. Baker,
and F. Kajzar, Phys. Rev. Lett. {\bf 62}, (1989) 1492; 
A.J. Heeger, D. Moses, and M. Sinclair, Synth. Met. {\bf 17}, (1987) 343.
\bibitem{su}J. Yu, B. Friedman, P.R. Baldwin, and W.P. Su, Phys. Rev. B 
{\bf 39}, (1989) 12814; J. Yu and W.P. Su, Phys. Rev. B {\bf 44}, (1991) 13315.
\bibitem{cwu}C.Q. Wu and X. Sun, Phys. Rev. B {\bf 42}, (1990) R9736.
\bibitem{shuai}Z. Shuai and J.L. Br\'{e}das, Phys. Rev. B {\bf 44}, (1991) 
R5962.
\bibitem{mazumdar}F. Guo, D. Guo, and S. Mazumdar, Phys. Rev. B {\bf 49}, 
(1994) 10102.
\bibitem{soos}F.C. Spano and Z.G. Soos, J. Chem. Phys. {\bf 99}, (1993) 9265.
\bibitem{mxu} Minzhong Xu and Xin Sun, to be submitted.
\bibitem{callaway}J. Callaway, Quantum Theory of the Solid State,
second edition (Academic Press, Inc, 1991), p.483.
\bibitem{weinreich} G. Weinreich, Solids: Elementary Theory for Advanced
Students (Wiley, New York, 1965), p.136.
\bibitem{heeger}A.J. Heeger, S. Kivelson, J.R. Schrieffer, and W.P. Su, Rev.
Mod. Phys. {\bf 60}, (1988) 781 and references there in.
\bibitem{mxu1}In the TLM model, using the static current operator 
$J_0=\sum_kA(k)\psi^\dagger_k\sigma_1\psi_k+B(k)\psi^\dagger_k\sigma_3\psi_k$,
where $A(k)=-ev_F\Delta/\varepsilon(k)$ and $B(k)=ev_F^2\hbar k/\varepsilon(k)$.
By $k \to k-eA/\hbar$, we have the IFC related to $\chi^{(1)}$: 
$J_A=\sum_kC(k)\psi^\dagger_k\sigma_3\psi_kA$, where 
$C(k)=-(ev_F\Delta)^2/\varepsilon^3(k)$. The computation of $J_A$ gives 
the first term in Eq. (\ref{x1}). But $J_A$ {\it cannot} be directly 
obtained from the field currents in the previous works,\cite{tlm,gebhard,wwu} 
thus it shows the difficulty to obtain IFCs in general models.  
\end{references}
\end{document}